\documentclass[a4paper]{jpconf}
\usepackage{latexsym,graphicx,amssymb,mathrsfs}
\usepackage{setspace,bm}
\usepackage{hyperref}

\usepackage[numbers,square,sort&compress]{natbib}
\graphicspath{{./Figures/}}

\newcommand{\diff}{\mathrm{d}}
\newcommand{\p}{\partial}
\newcommand{\ve}{\varepsilon}

\newcommand{\up}{\uparrow}
\newcommand{\down}{\downarrow}
\newcommand{\be}{\begin{equation}}      
\newcommand{\ee}{\end{equation}}      
\newcommand{\bea}{\begin{eqnarray}}      
\newcommand{\eea}{\end{eqnarray}}

\begin{document} 

\title{Functional renormalization group approach to conventional theory of superfluidity and beyond}

\author{Yuya Tanizaki,$^{1,2}$ Gergely Fej\H{o}s,$^2$ and Tetsuo Hatsuda$^{2,3}$}

\address{$^1$Department of Physics, The University of Tokyo, Tokyo 113-0033, Japan}
\address{$^2$Theoretical Research Division, Nishina Center, RIKEN, Wako 351-0198, Japan}
\address{$^3$Kavli IPMU (WPI), The University of Tokyo, Chiba 277-8583, Japan}
\ead{yuya.tanizaki@riken.jp}

\begin{abstract}
Fermionic functional renormalization group (FRG) is applied to describe the superfluid phase transition of the two-component fermionic system with attractive contact interaction. Connection between the fermionic FRG approach and the Bardeen-Cooper-Schrieffer (BCS) theory with its Gorkov and Melik-Barkhudarov (GMB)  correction is made clear, and the FRG flow of the fermion self-energy is also studied to go beyond the BCS+GMB theory. 
The superfluid transition temperature and the associated chemical potential are calculated in the region of the negative scattering length using fermionic FRG.
\end{abstract}


\section{Introduction\label{intro}}

 Superfluidity in  many-fermion systems is one of the central problems in condensed matter, atomic, nuclear and particle physics. Examples include liquid superfluid $^3$He, electron superconductivity, cold atoms, nucleon superfluidity, color superconductivity, etc. \cite{Cooper_Feldman201006}.  
We report our recent study on the application of fermionic functional renormalization group (f-FRG) method to the superfluid phase transition of two-component fermionic systems \cite{Tanizaki:2013doa}. The purpose of this study is to develop a series of approximations without introducing the auxiliary bosonic field to make a firm connection between the non-perturbative FRG approach and the conventional Bardeen-Cooper-Schrieffer (BCS) theory with its Gorkov and Melik-Barkhudarov (GMB) correction. 
Furthermore, we wish to explore the role of the RG flow of the fermion self-energy to go beyond BCS+GMB theory.
We note that such analyses can be best achieved by fermionic FRG without bosonization, because f-FRG can provide a systematic and unbiased study of interacting fermions \cite{shankar1994renormalization,RevModPhys.84.299}.

\section{Fermionic FRG Formalism\label{form}}
 We consider non-relativistic two-component fermions $\psi=(\psi_{\up},\psi_{\down})$ with a contact interaction:
\begin{equation}
S[\overline{\psi},\psi]=\int_0^{\beta}\diff \tau\int \diff^3\bm{x}
\left[\overline{\psi}\left(\p_{\tau}-{\nabla^2\over 2m}-\mu\right)\psi
+g\overline{\psi}_{\up}\overline{\psi}_{\down}\psi_{\down}\psi_{\up}\right],
\label{intro02}
\end{equation}
where $\beta(=1/T)$, $\mu$, $m$ and $g$ are the inverse temperature, the chemical potential,
 the mass and the bare coupling constant, respectively. 
The classical action (\ref{intro02}) is symmetric under global $U(1)$ 
phase rotation, $SU(2)$ spin rotation, and space-time translation. 

Following the idea of 
 the FRG method \cite{wetterich1993exact}, we define a scale-dependent average effective action $\Gamma_k[\overline{\psi},\psi]$ as a one-particle-irreducible (1PI) effective action of $S[\overline{\psi},\psi]+\overline{\psi}R_k\psi$, where  
$R_k$ is an infrared (IR) regulating function. 
We consider the following Ansatz of $\Gamma_{k}$:
\bea\fl\quad
\Gamma_k[\overline{\psi},\psi]
&=&\int_p^{(T)}\overline{\psi}(p) [G^{-1}(p)-\Sigma_k(p)] \psi(p)\nonumber\\
\fl&+&\int_{p,q,q'}^{(T)}\Gamma_k^{(4)}(p; q,q')\overline{\psi}_{\up}({p\over2}+q)\overline{\psi}_{\down}({p\over2}-q)\psi_{\down}({p\over2}-q')\psi_{\up}({p\over2}+q'),
\label{form06}
\eea
where  $G^{-1}(p)=ip^0+{\bm{p}^2\over 2m}-\mu$ is the inverse propagator with $p=(p^0,\bm{p})$, $\Sigma_k(p)$ is the self-energy, and $\Gamma_k^{(4)}(p;q,q')$ is the four-point vertex.  Also, we adopt an abbreviated notation,   $\int_p^{(T)}\equiv \int{\diff^3\bm{p}\over (2\pi)^3}T\sum_{p^0}$. 
In this study, we neglect the momentum dependence of the self-energy and define its constant part by $\sigma_k\equiv \Re \Sigma_k(\pm\pi T,\bm{0})$. 
IR regulator $R_k(\bm{p})$ is chosen to suppress one-particle excitations around the Fermi surface with the excitation energy smaller than $k^2/2m$: $R_k(\bm{p})=\Big[{k^2\over 2m}\mathrm{sgn}(\xi(\bm{p}))-\xi(\bm{p})\Big]\theta\Big({k^2\over 2m}-|\xi(\bm{p})|\Big)$ with $\xi(\bm{p})=\bm{p}^2/2m-\mu-\sigma_0$.  
$\Gamma_k$ obeys the flow equation 
\be
\p_k \Gamma_k = {1\over 2}\int_p^{(T)}\mathrm{Tr}\left[{\p_kR_k(p)\over \Gamma_k^{(2)}(p)+R_k(p)}\right]. 
\label{eq3}
\ee

Our primary goal is to calculate the ratios $T_c/\ve_F$ and $\mu/\ve_F$ 
as a function of the dimensionless constant $1/(k_{\rm F} a_{\rm s})$:
Here  $a_{\rm s}$ is the scattering length between fermions, $k_{\rm F}\equiv(3\pi^2 n)^{1/3}$ with
  $n$ being the fermion number density and  $\ve_F\equiv k_F^2/2m$. For a second-order phase transition, the critical point $T=T_c$ can be determined from the Thouless criterion \cite{thouless1960perturbation} by looking at the divergence  of the   fermion-fermion scattering matrix at the total momentum  $p=0$:\be
\left[ \Gamma_{k=0}^{(4)}(p=0) \right]^{-1} =0 \ \ {\rm at}\ T=T_c.  
\label{thouless_cr}
\ee
Note that  the number density $n$ is related to $T$ and $\mu$ through the number equation:
\begin{equation}
n=\langle \overline{\psi}\psi \rangle = 2\int^{(T)}_p {-1\over G^{-1}(p)-\sigma_{0}}. 
\label{number_eq}
\end{equation}

\section{Flow equation of f-FRG for self-energy and four-point vertex}\label{sec:Flow_Eq}

\begin{figure}[t]
\centering
$\partial_k \parbox{3em}{\includegraphics[width=3em]{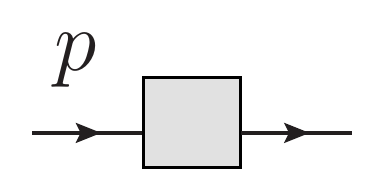}}
=\widetilde{\partial}_k \parbox{3.5em}{\includegraphics[width=3.5em]{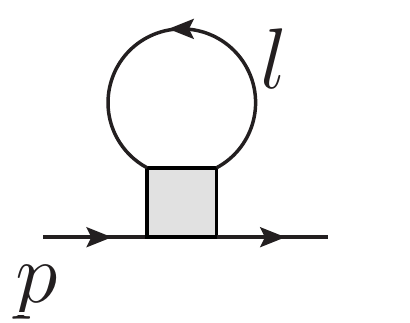}}$ \quad
$\displaystyle \partial_k \parbox{3.8em}{\includegraphics[width=3.8em]{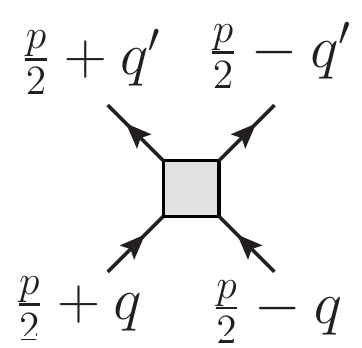}}
=\widetilde{\partial}_k\Biggl(
\parbox{3.6em}{\includegraphics[width=3.6em]{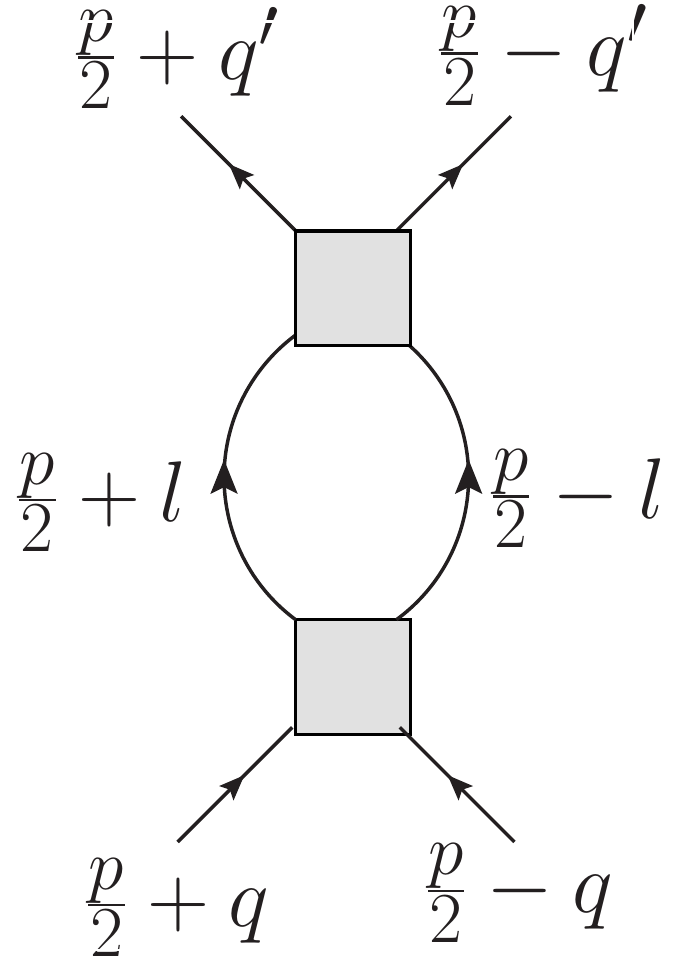}}+\sum_\pm
\parbox{6.5em}{\includegraphics[width=6.5em]{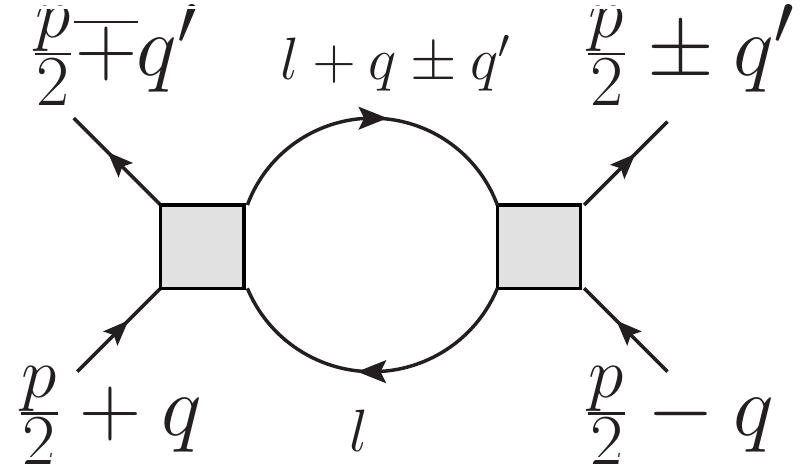}}
\Biggr)$
\caption{Flow equation of the self energy $\Sigma_k$ and the four-point vertex $\Gamma_k^{(4)}$. 
\label{fig:flow_4pt}}
\end{figure}

Taking the vertex expansion (\ref{form06}) of the scale dependent 1PI effective action  $\Gamma_k[\overline{\psi},\psi]$, 
 a closed set of equations for the self-energy and four-point vertex can be derived from (\ref{eq3}). They are given diagrammatically in Fig. \ref{fig:flow_4pt}. 
In order to simplify the flow equation of four-point vertex, we neglect its momentum dependence and regard it as an effective coupling constant. At the lowest momenta, our flow equations can be approximated as
\bea
\label{eq:coup_sigma-flow}
\fl\qquad\quad\partial_k \sigma_k&=&\Re\widetilde{\partial}_k \int_l^{(T)} \frac{\Gamma_k^{(4)}(\pm\pi T+l^0,{\bm{l}})}{
G^{-1}(l)-\sigma_k+R_k(l)}, \\
\label{eq:coup_gamma-flow}
\fl\quad\partial_k \Gamma_k^{(4)-1}(0)&=&\widetilde{\partial}_k \int_l^{(T)} \frac{1}{[G^{-1}(l)
-\sigma_k+R_k(l)][G^{-1}(-l)-\sigma_k+R_k(-l)]}  \nonumber\\
\fl&+&\widetilde{\partial}_k \int\limits_{0}^{8m\mu}\frac{\diff |\bm{Q}|^2}{8m\mu} \int_l^{(T)} \frac{1}{[G^{-1}-\sigma_k+R_k](l)[G^{-1}-\sigma_k+R_k](Q+l)},
\eea
where transfer momenta $Q=(0,\bm{Q})$. Here, $\widetilde{\p}_k$ is the derivative of $k$ acting only on the IR regulator $R_k$. 
On the r.h.s. of (\ref{eq:coup_gamma-flow}), the first and second terms refer to the particle-particle (PP) and particle-hole (PH) fluctuations, respectively. 
For the PH loop contribution of the flow of four-point coupling, the relative momenta is projected onto the Fermi surface. 
As already mentioned, in (\ref{eq:coup_sigma-flow}), the frequency of the fermion self-energy is restricted to the lowest values
$\pm \pi T$. 
After performing the Matsubara sum, $\sigma_k$ and $\sigma_0$ appear as a combination $\sigma_0-\sigma_k$.
Since the approximate particle-hole symmetry would make  this combination small for $k < k_F$, where PH contribution will turn out to be already not significant, we take $\sigma_k=\sigma_0$ in the
actual calculation of this contribution.
  
To take into account the momentum dependence in a minimal way, we make the following
expansion of the four-point vertex in terms of center-of-mass momentum and keep first few terms:
\bea
\label{Eq:expan}\hspace{-3em}
\Gamma_k^{(4)-1}(p^0+l^0,{\bm{l}})
 \approx -Z_k^{-1} \left[ i(l^0 + p^0)+ S_k^{(1)} \cdot |{\bm{l}}| + S_k^{(2)}\cdot {\bm{l}}^2+|\mu^B_k|\right] ,\eea
where $ |\mu_k^B|=-Z_k \Gamma_k^{(4)-1}(0)$. Here we adopt a hybrid approach in which 
 $\Gamma_k^{-1}(0)$ is calculated from the flow equation, while its derivatives $Z_k,S_k^{(1,2)}$ with respect
 to the frequency and momentum are estimated by the solution of an approximated flow equation keeping only the PP-contribution (this actually coincides with the usual Random Phase Approximation). 
  After applying the expansion (\ref{Eq:expan}), the flow equation of the self-energy becomes
\bea
\label{Eq:flow_sigma_a}
\fl\quad\partial_k \sigma_k&=&\Re\tilde{\partial}_k \int_l^{(T)} \frac{1}
{G^{-1}(l)-\sigma_k+R_k(l)}\frac{-Z_k}{i(l^0 \pm \pi T)+S_k^{(1)}|\bm{l}|+S_k^2\bm{l}^2+|\mu_k^B|} ,\nonumber\\
\label{Eq:flow_sigma_b}
\fl&=&-\Re\tilde{\partial}_k\int_0^{\infty}\frac{dll^2}{2\pi^2}\Big(n_B(\omega_k^B(\bm{l}))+n_F(\omega_k(\bm{l}))\Big)
\frac{Z_k}{\omega_k(\bm{l})-\omega_k^B(\bm{l})\mp i \pi T},
\eea
where $\omega_k(\bm{l})=\frac{\bm{l}^2}{2m}-\mu-\sigma_k+R_k({\bm{l}})$, and $\omega_k^B(\bm{l})=|\mu_k^B|+S_k^{(1)}|{\bm l}|+S_k^{(2)}{\bm l}^2$, with $n_B$ being the Bose distribution function and $n_F$ the Fermi distribution function. 

\section{Results}\label{sec:results}
Let us briefly discuss the connection of f-FRG and the BCS+GMB theory. If we take into account only the PP correlation in the flow equation with neglecting the self energy (i.e. $\sigma_k\equiv0$) and the PH loop in the four-point coupling flow, in the weakly coupled regime we reproduce the BCS transition temperature by applying the Thouless criterion: $T_c^{\mathrm{BCS}}/\ve_F=0.613 \exp(-\pi/(2 k_F|a_s|))$. By taking into account the PH loop contribution (evaluating it at zero temperature) and still neglecting the self-energy, the transition temperature is reduced by a factor $(4e)^{1/3}\approx 2.2$, which is nothing but the GMB correction of the critical temperature: $T_c^{\mathrm{GMB}}=T_c^{\mathrm{BCS}}/2.2$.

\begin{figure}[t]
\begin{minipage}{0.49\hsize}\centering\hspace{-1.8em}
\includegraphics[scale=0.3]{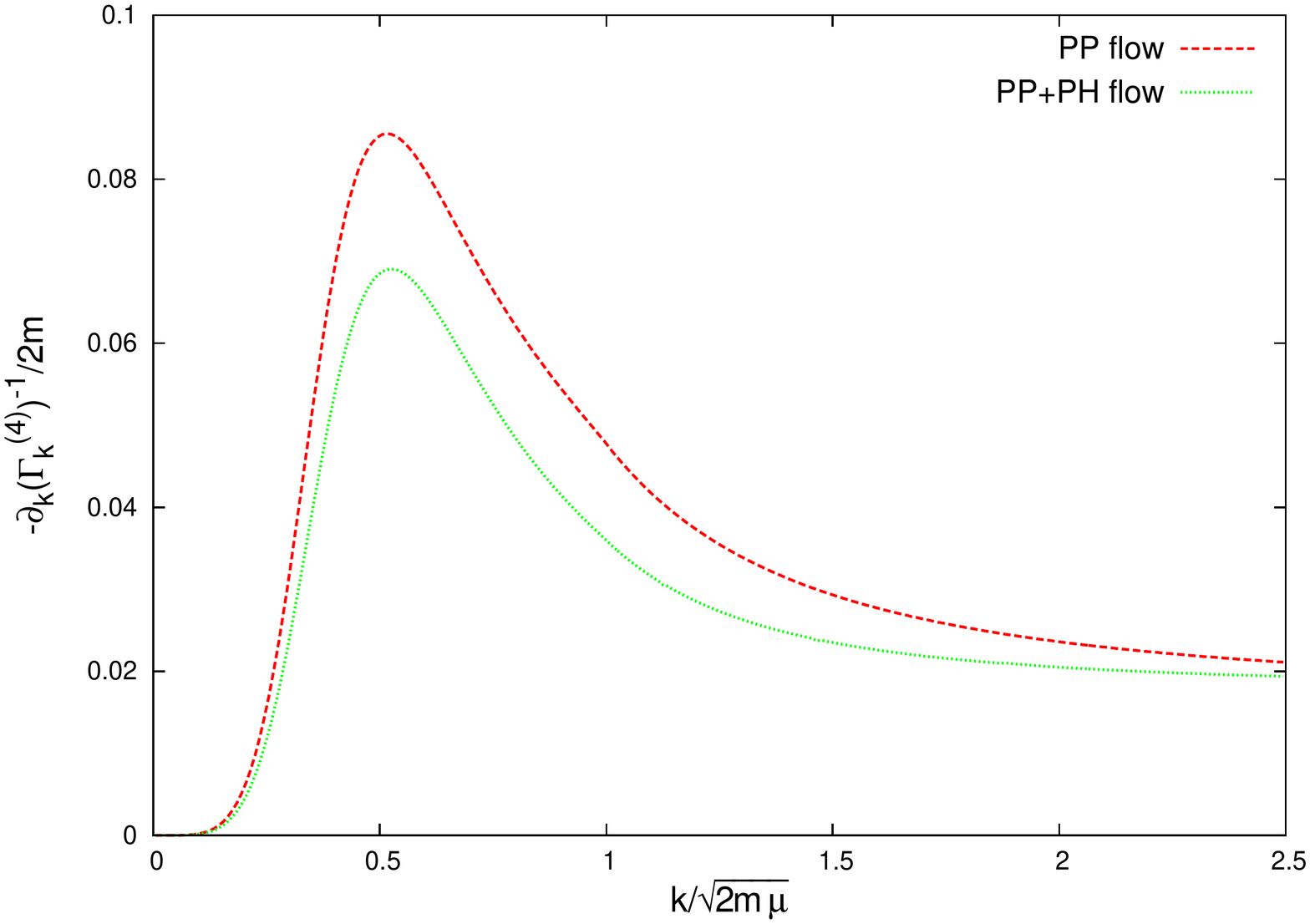} \vskip -20pt
\caption{Derivative of $1/\Gamma_k^{(4)}(0)$ at $T/\mu=0.06$ with and without the PH loop included.}
\label{Fig3}
\end{minipage}\hspace{0.02\hsize}
\begin{minipage}{0.49\hsize}\centering\hspace{-1.8em}
\includegraphics[scale=0.3]{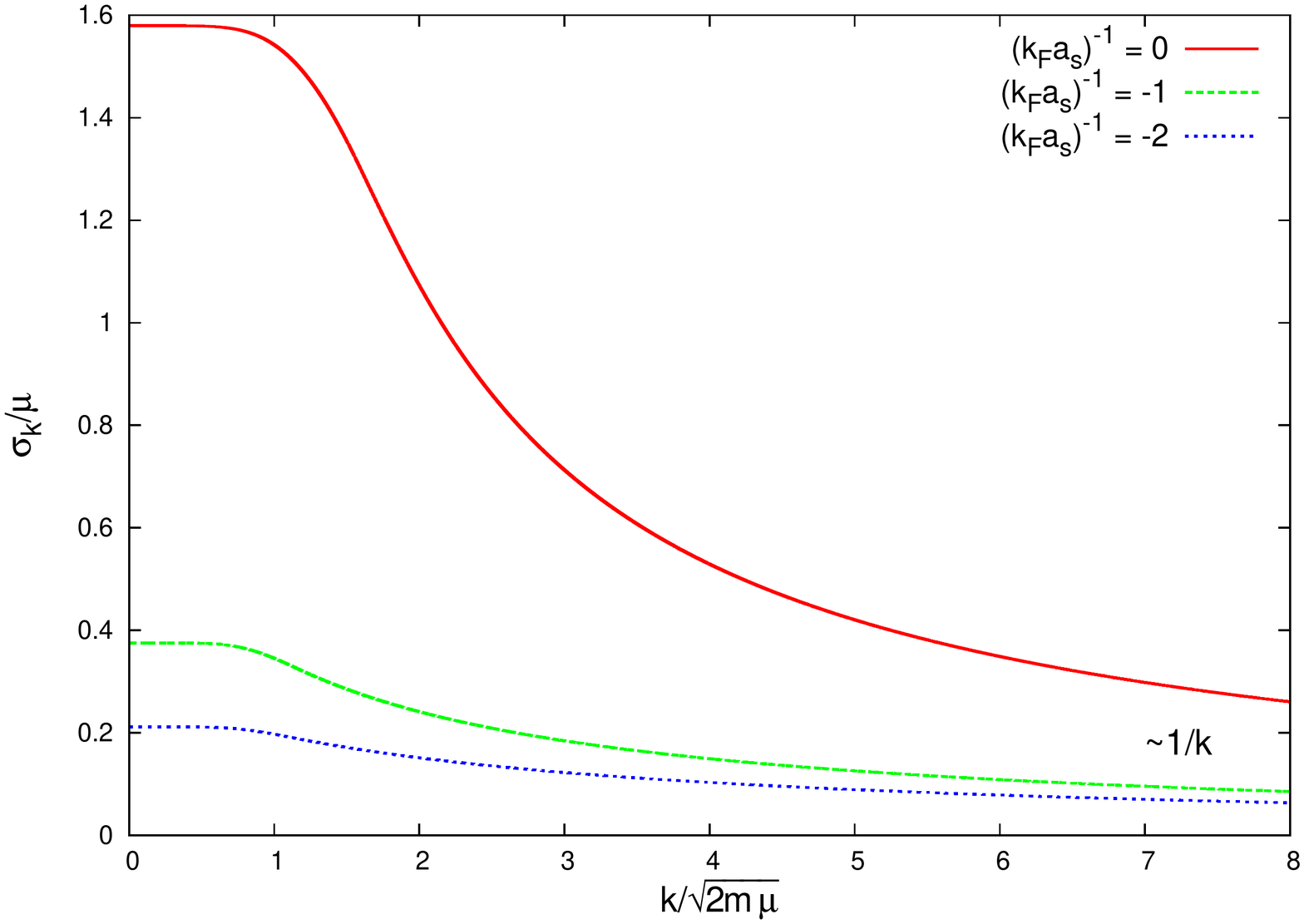} \vskip -20pt
\caption{Self-energy $\sigma_k$ as a function of $k$ for $(k_F a_s)^{-1}=-2,-1,0$ at $T=T_c$.}
\label{Fig4}
\end{minipage}
\end{figure}

\begin{figure}
\begin{minipage}{0.49\hsize}\centering\hspace{-1.8em}
\includegraphics[scale=0.3]{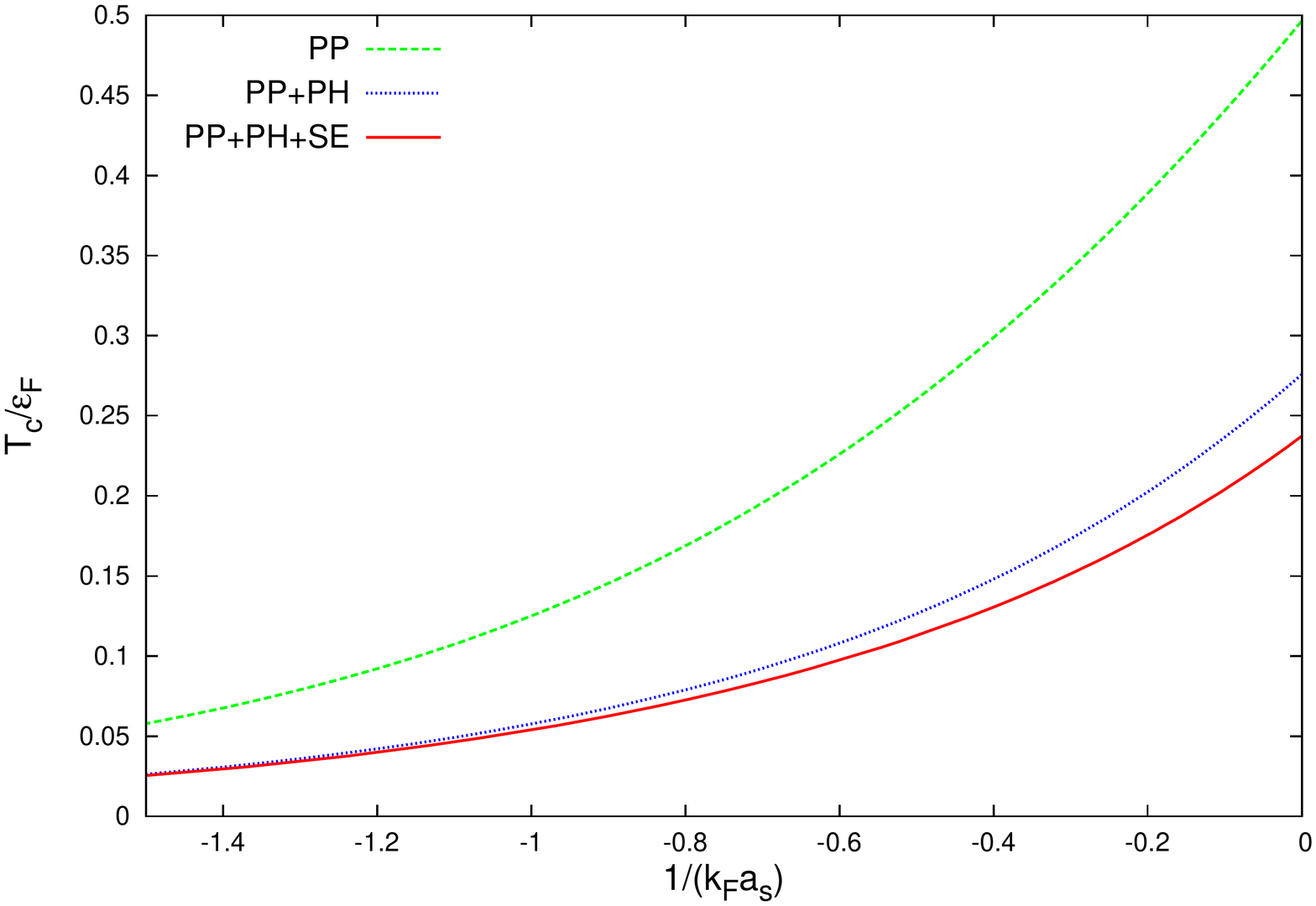}\vskip -20pt
\caption{ $T_c/\ve_F$ as a function of $(k_F a_s)^{-1}$ 
in three different levels of approximations.
} 
\label{Fig5}
\end{minipage}\hspace{0.02\hsize}
\begin{minipage}{0.49\hsize}\centering\hspace{-1.8em}
\includegraphics[scale=0.3]{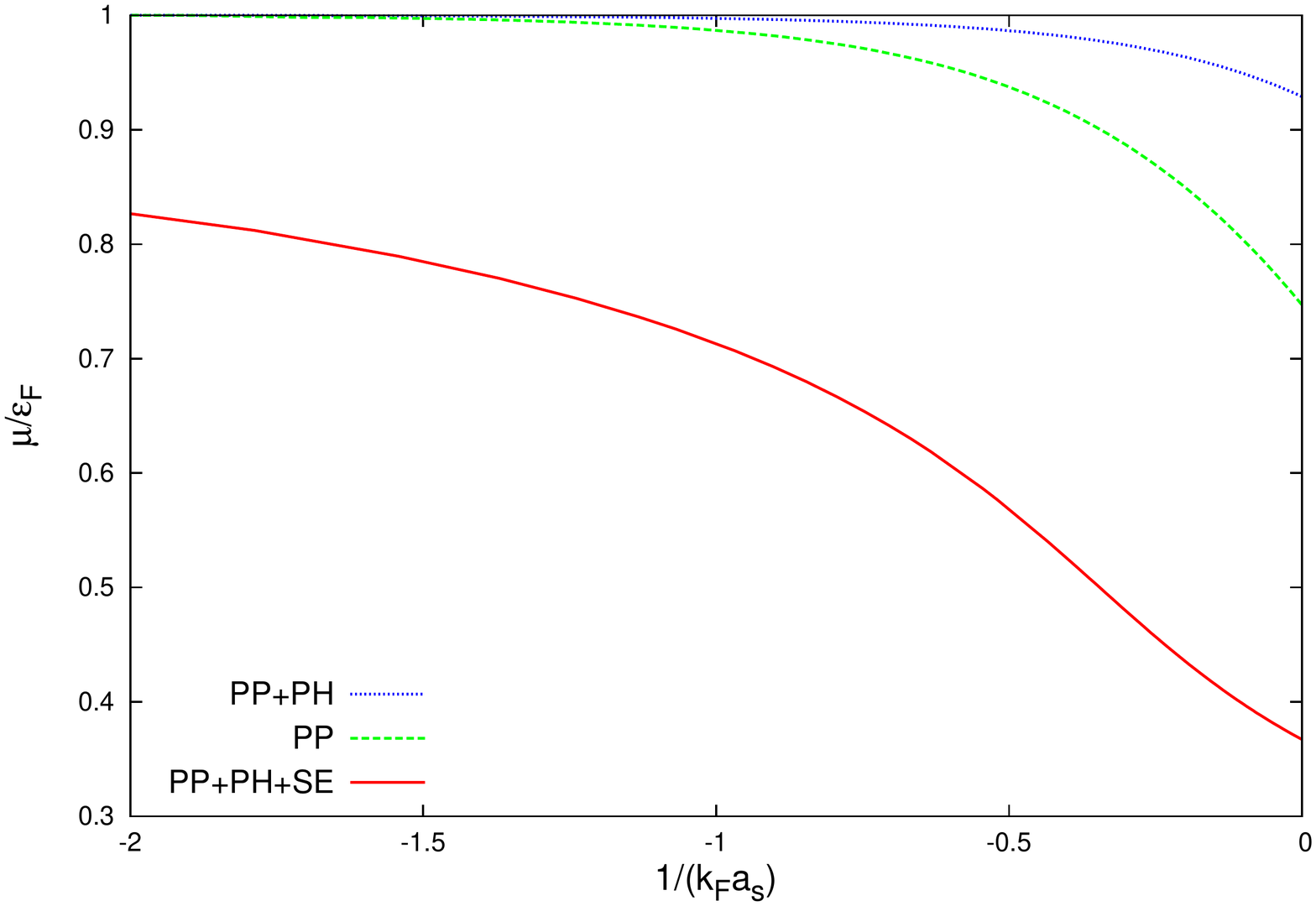}\vskip -20pt
\caption{$\mu/\ve_F$ as a function of $(k_F a_s)^{-1}$.
The meaning of each line is the same as Fig.\ref{Fig5}. 
}
\label{Fig6}
\end{minipage}
\end{figure}

On Fig.\ref{Fig3}, the flow of $-\p_k(1/\Gamma_k^{(4)}(0))$ is plotted with and without particle-hole (PH) loop when the self-energy correction is neglected. The flows of the self-energy for $(k_F a_s)^{-1}= -2, -1, 0$ at $T_c$  are shown in  Fig.\ref{Fig4}.
  Fig.\ref{Fig5} and Fig.\ref{Fig6} show our numerical results of the critical temperature ($T_c/\varepsilon_F$) and the chemical potential ($\mu/\varepsilon_F$)  as a function of the inverse scattering length ($(k_F a_s)^{-1}<0$), respectively, for three different levels of approximations: PP contains only the effect of particle-particle correlation, PP+PH just neglects the self-energy, and PP+PH+SE is our final result including self-energy (SE) as well.  
    
We note that Fig.\ref{Fig3} provides a physical interpretation of GMB correction of the BCS theory from a renormalization group point of view. When the self-energy is neglected, but the contribution of PH loop is included in the flow of four-point coupling, $\p_k(1/\Gamma_k^{(4)}(0))$ is reduced only around the region $k\sim k_F$. 
Since the particle-particle loop contribution reproduces exactly the BCS result as the low-energy effective theory, this implies that the effective coupling of the BCS theory is weakened due to the existence of matter, and such screening occurs around intermediate energy scales $k^2/2m\sim \ve_F$. 

Considering the self-energy equation for large $k$ $(\gtrsim k_F)$, it can be shown that its flow can be well described by perturbation theory and that it behaves as $\sim 1/k$ (see Fig. \ref{Fig4}).
 On the other hand, it almost stops for small $k$ $(\lesssim k_F)$: This is due to the fact that the approximate particle-hole symmetry valid for small $k$ protects the shift of the Fermi level. 
 Fig. \ref{Fig4} shows that the magnitude of the self-energy $\sigma_0$ becomes larger as the coupling becomes strong, while the saturation of $\sigma_k$ at $k\sim k_F$ takes place for any coupling strength.

As can be seen from Fig. \ref{Fig5},  
PH correlation is the dominant source of the reduction of $T_c/\ve_F$, and, as already discussed, its physical origin is the screening of the coupling strength. We should note that this does not imply the smallness of the self-energy correction. Since the flow of self-energy stops for $k\lesssim k_F$, it does not affect low-energy dynamics after shifting the Fermi level away from the chemical potential. 
The self-energy correction still leads to further small reduction of $T_c/\ve_F$, since it makes $\ve_F$ (or the number density $n$) bigger, as the coupling strength becomes stronger towards the unitary regime, but the effect is 
rather limited because $\sigma_0-\sigma_k$ around $k\sim k_F$ cannot be large due to the particle-hole symmetry, as discussed above. 

Fig. \ref{Fig6} shows that, the effects of the PH correlation to $\mu/\ve_F$ are opposite of what we find for $T_c/\ve_F$. This is because the critical temperature $T_c$ of PP approximation is higher than that of the PP+PH result for the same chemical potential $\mu$. Therefore, the number of fermions in the PP case is larger than that of PP+PH. 
When the self-energy correction is taken into account, the number density $n$ and hence the Fermi energy $\varepsilon_F$ increase for given $\mu$, therefore a considerable decrease of $\mu/\ve_F$ arises.

 Although it is beyond our scope to predict a quantitatively correct $T_c$ in the unitary regime, 
 it is still instructive to compare our result at unitarity, $(T_c/\ve_F,\mu/\ve_F)=(0.237, 0.367)$,
 with the previous FRG results using auxiliary bosonic field,
 $(T_c/\ve_F, \mu/\ve_F)\simeq (0.248, 0.51)$ \cite{PhysRevA.81.063619}. 
For convenience of readers, we also quote some results of Monte Carlo simulations for $(T_c/\ve_F,\mu/\ve_F)$:  $(0.152(7),0.493(14))$  \cite{PhysRevLett.96.160402,PhysRevLett.101.090402}, $(0.15(1),0.43(1))$ \cite{PhysRevA.78.023625}, $(0.171(5),0.429(9))$ \cite{PhysRevA.82.053621}. 

\section{Summary}\label{sec:conclusions}

{
We developed a fermionic FRG to describe the superfluid phase transition for two-component fermions with a contact interaction.
 By making vertex expansion of the 1PI effective action $\Gamma_k[\bar{\psi},\psi]$ up to the four-point vertex and solving RG flow equations, we determined the critical temperature $T_c$ in the regime of negative scattering lengths. 
In order to clarify the relation between  the FRG approach and the BCS+GMB theory and to go beyond,
 we have taken into account the PP and PH correlations together with the self-energy correction, and the role of each is analyzed in details in their respective flow equation. 

  Resultant value of  $T_c/\ve_F$  does not receive
   large correction from the self-energy except for the unitary regime ($1/(k_Fa_s) \rightarrow 0$).
 On the other hand, $\mu/\ve_F$ shows a large reduction  by the self-energy correction due to the increase of the Fermi energy given $\mu$. It shows that the self-energy correction is still comparable with the chemical potential even for $(k_F a_s)^{-1}\lesssim -1$, however, its effect on $T_c/\ve_F$ is almost negligible.  Extrapolation to unitarity gives $T_c/\ve_F=0.237$ and $\mu/\ve_F=0.367$.

\section*{Acknowledgements}
Y. T. is supported by JSPS Research Fellowships for Young Scientists. 
G. F. is supported by the Foreign Postdoctoral Research program of RIKEN. 
T. H. is partially supported by RIKEN iTHES project.

\bibliography{FRG,BCS_BEC,fFRG}
\end{document}